\documentclass[aps,prm,twocolumn,superscriptaddress,floatfix,10pt]{revtex4}
\usepackage{graphicx}
\usepackage{amsmath}
\usepackage{color}
\begin{document}

\title{Elasticity and thermal transport of commodity plastics}

\author{C\'eline Ruscher}
\affiliation{Stewart Blusson Quantum Matter Institute, University of British Columbia, Vancouver BC V6T 1Z4, Canada}
\affiliation{Department of Physics and Astronomy, University of British Columbia, Vancouver BC V6T 1Z1, Canada}
\author{J\"org Rottler}
\affiliation{Stewart Blusson Quantum Matter Institute, University of British Columbia, Vancouver BC V6T 1Z4, Canada}
\affiliation{Department of Physics and Astronomy, University of British Columbia, Vancouver BC V6T 1Z1, Canada}
\author{Charlotte Boott}
\affiliation{Department of Chemistry, University of British Columbia, Vancouver BC V6T 1Z1, Canada}
\author{Mark J. MacLachlan}
\affiliation{Stewart Blusson Quantum Matter Institute, University of British Columbia, Vancouver BC V6T 1Z4, Canada}
\affiliation{Department of Chemistry, University of British Columbia, Vancouver BC V6T 1Z1, Canada}
\author{Debashish Mukherji}
\email[]{debashish.mukherji@ubc.ca}
\affiliation{Stewart Blusson Quantum Matter Institute, University of British Columbia, Vancouver BC V6T 1Z4, Canada}

\date{\today}

\begin{abstract}
	Applications of commodity polymers are often hindered by their low thermal conductivity. 
	In these systems, going from the standard polymers dictated by weak van der Waals 
	interactions to biocompatible hydrogen bonded smart polymers, the thermal transport coefficient 
	$\kappa$ varies between $0.1-0.4$ Wm$^{-1}$K$^{-1}$. Combining all-atom molecular dynamics simulations 
	with some experiments, we study thermal transport and its link to the elastic response of (standard and smart) 
	commodity plastics. We find that there exists a maximum attainable stiffness (or sound wave velocity), thus providing an
	upper bound of $\kappa$ for these solid polymers. The specific chemical structure and the glass transition 
	temperature play no role in controlling $\kappa$, especially when the microscopic interactions are
	hydrogen bonding based. Our results are consistent with the minimum thermal conductivity model and 
	existing experiments. The effect of polymer stretching on $\kappa$ is also discussed.
\end{abstract}

\maketitle

\section{Introduction}
\label{sec:intro}

In crystals, the periodicity of the crystal lattice allows for the propagation of vibrational excitations 
(or phonons) that carry a heat current. Here, the coefficient of thermal conductivity $\kappa$ is directly related to the heat capacity $c$, 
the phonon mean-free path $\Lambda$, and the group velocity $v_{\rm g}$ \cite{cahill03jap,fuga17ps}. 
Most commonly known crystals have $\kappa \gtrsim 100$ Wm$^{-1}$K$^{-1}$ \cite{slack73jpcs},
which can even exceed $1000$ Wm$^{-1}$K$^{-1}$ in carbon-based materials \cite{tomanek00prl,davide15prb,chang17prl,zhang17prl}. 
On the other hand, $\Lambda \to 0$ for disordered (amorphous) materials, leading to $\kappa < 10$ Wm$^{-1}$K$^{-1}$ \cite{cahill92prb}. 
These materials are broadly classified as hard matter having large cohesive energy densities. 
Moreover, these systems often require rather cumbersome materials processing, can be 
expensive and often have restrictive flexibility. In this context, another class 
consists of polymers that also form amorphous state. 
Broadly speaking, polymers are classified as soft matter where the relevant energy scale is 
comparable to the thermal energy $k_{\rm B}T$ with temperature $T = 300$ K 
and $k_{\rm B}$ being the Boltzmann constant \cite{doibook}. Therefore, the properties of polymers are dictated by 
large fluctuations, thus are high flexible, easily processable and most times inexpensive.

Amorphous polymers are materials that provide a suitable platform for the 
flexible design of advanced functional materials \cite{cohen10natmat,mukherji14natcom,sissi14natcom,winnik15review,hoogenboom,mukherji17natcom}. 
Moreover, solvent free (dry) states of common polymers, often referred to as commodity plastics, 
usually have rather small $\kappa < 0.5$ Wm$^{-1}$K$^{-1}$ \cite{choy77pol,shenogin09jap,pipe14nm,cahill16mac,luo16pccp,cahill17prb,mukherji19mac}. 
While small $\kappa$ values are extremely desirable for thermoelectric materials \cite{ruan09prb,shuai17afm}, they also 
create severe problems when used under high temperature conditions for electronic packaging, organic solar cells, 
and organic light emitting diodes, to name a few \cite{kim05apl,cola16aami}. 

The properties of commodity plastics, such as polyethylene (PE), polystyrene (PS), and/or poly(methyl methacrylate) (PMMA), 
are dictated by weak van der Waals (vdW) interactions resulting in $\kappa < 0.2$ Wm$^{-1}$K$^{-1}$ \cite{choy77pol,cahill16mac} 
These polymers are water insoluble because of their dominant carbon-based (hydrophobic) architectures. 
The bonded interactions in these polymers are based on carbon-carbon covalent bonds, 
having a bond strength of $80~k_{\rm B}T$ \cite{polh} and thus live forever under the 
unperturbed conditions, creating severe environmental problems. Therefore, the recent interest has been
directed towards peptide-based polymers that are bio-compatible/bio-degradable, and are often referred to as
commodity ``smart" polymers. These systems are also water soluble because of their dominant hydrogen 
bonded (H$-$bond) nature \cite{cohen10natmat,winnik15review,hoogenboom}. The dry states of these H$-$bond based polymers 
usually have $\kappa \sim 0.3-0.4$ Wm$^{-1}$K$^{-1}$ \cite{pipe14nm,cahill16mac,mukherji19mac}. 

The thermal conductivity of these commodity plastics has a rather restrictive tunability limiting their 
broad applications. Traditionally, extensive efforts have been devoted to tune $\kappa$ of polymers by 
nanomaterials blending \cite{cola16aami,keb11jap,good17natmat,bock16pol}. 
Other strategies include polymer blending, cross linking, and 
macromolecular engineering \cite{pipe14nm,cahill16mac,mukherji19mac,zishun18aami}.
Advances are usually carried from the experimental side with several carefully conducted works,
while simulation efforts are rather limited \cite{shenogin09jap,luo16pccp,mukherji19mac,zishun18aami,chen08prl,yang12prb,luo13acsnano,chen19jap}.
The majority of these simulation efforts deal with vdW-based polymers.

\begin{figure*}[ptb]
\includegraphics[width=0.94\textwidth,angle=0]{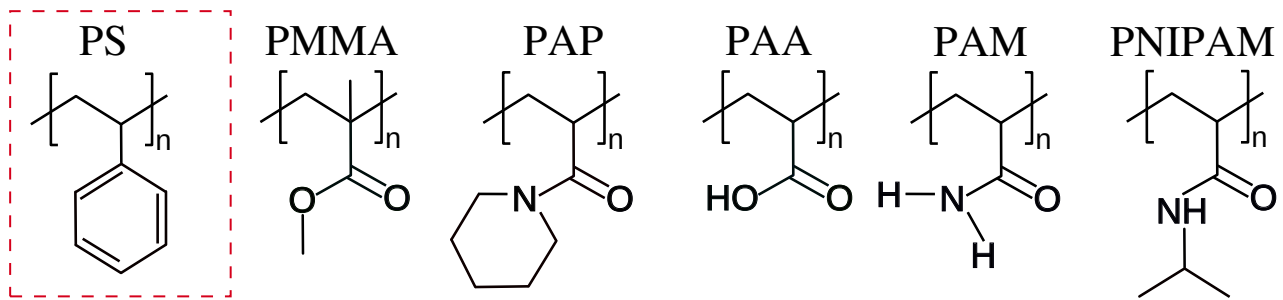}
	\caption{Schematics showing different monomeric structures investigated in this study: poly(styrene) (PS), 
	poly(methyl methacrylate) (PMMA), poly({\it N}-acryloyl piperidine) (PAP), poly(acrylic acid) (PAA), 
	poly(acrylamide) (PAM), poly({\it N}-isopropyl acrylamide) (PNIPAM) and copolymer P(AM-co-NIPAM) at 
	20\% AM monomer mole fraction. We have simulated chains of all these systems except for PS. 
	For PS we discuss experimental and simulation data from the published literature only.
\label{fig:schem}}
\end{figure*}

In this work, we use large-scale all-atom molecular dynamics simulations to relate 
microscopic molecular organization, propensity of forming hydrogen bonds, mechanical response 
and its links to the thermal transport of solid macromolecular systems. 
For this purpose, we have chosen a set of six different linear (co-)polymers in their solvent free (dry) states: 
namely PMMA, poly({\it N}-acryloyl piperidine) (PAP), poly(acrylic acid) (PAA), 
polyacrylamide (PAM), poly({\it N}-isopropyl acrylamide) (PNIPAM) and copolymer P(AM-co-NIPAM) at 20\% AM monomer 
mole fraction. The schematic representation of monomer structures of all these polymers are shown in Fig. \ref{fig:schem}.
We have chosen these particular polymeric systems because of their experimental relevance \cite{cahill16mac}.
Furthermore, while the properties of PAP and PMMA are dictated by pure vdW interactions and for PAM and PAA pure H$-$bonds are
dominant, we have also chosen two additional systems, namely PNIPAM and P(AM-co-NIPAM), where the delicate balance between
H-bonds and vdW interactions plays a key role (see Section IIIB for more details).
For comparison, we have also re-analyzed simulation data of H$-$bonded assymetric PAA-PAM blends 
obtained in our earlier work \cite{mukherji19mac}. 
Experiments are also performed for pure PNIPAM, where experimental data 
was not available. Our analysis suggests that there is a maximum in attainable $\kappa$ 
for commodity plastics, which can be attributed to the maximum in materials stiffness and is 
consistent with the minimum thermal conductivity model \cite{cahill92prb}.

The remainder of the paper is organized as follows: In Section \ref{sec:method}, we sketch our methodology. Results and discussions 
are presented in Section \ref{sec:res} and finally we draw our conclusions in Section \ref{sec:conc}.

\section{Method, model and materials}
\label{sec:method}

\subsection{Molecular dynamics simulations}

The GROMACS molecular dynamics package is used for the initial equilibration of different polymers in their (solvent free) 
melt states at $T = 600$ K \cite{gro}. These equilibrated samples are subsequently quenched down to $T = 300$ K where the 
thermal transport calculations are performed using the LAMMPS molecular dynamics package \cite{lammps}.

The initial equilibration in GROMACS is performed in the isobaric ensemble, where the temperature is
imposed using the Berendsen thermostat with a coupling constant of 2 ps and the pressure is set 
to 1 bar with a Berendsen barostat with a coupling time of 0.5 ps \cite{berend}. Electrostatics are treated using the particle-mesh Ewald method \cite{pme}. 
The interaction cutoff for non-bonded interactions is chosen as 1.0 nm. The simulation time step is chosen as $\triangle t = 1$ fs
and the equations of motion are integrated using the leap-frog algorithm \cite{lfa}. All bond vibrations are constrained using the LINCS algorithm \cite{linc}.

A polymer chain length of $N_{\ell} = 30$ monomers was chosen for PMMA, PAP, PAA and PAM, while $N_{\ell} = 32$ is taken for PNIPAM and P(NIPAM-co-AM). 
Note that the latter two systems have slightly larger $N_{\ell}$ because these configurations are taken from our 
earlier works \cite{mukherji17natcom,mukherji19mac,oliveira17jcp}. These chains are typically $12-15~\ell_{p}$, with $\ell_{p}$ being the persistence length.
Chains are placed randomly in a cubic simulation box where the equilibrated linear dimensions $L$ vary between 7.7$-$8.3 nm.
Simulations are performed for 80 ns at $T=600$ K. This time is around one order of magnitude larger 
than the longest relation time for all polymers investigated herein.

The standard OPLS force field \cite{OPLS96} is used to simulate all polymers except for PMMA and PAM. 
For PMMA \cite{mukherji17natcom} and PAM \cite{oliveira17jcp}, we have taken the modified force field 
parameters used earlier to study single chain conformations in aqueous solutions. 
To ensure if these force field parameters can also capture properties of solid polymeric materials, we have calculated 
the glass transition temperatures $T_{\rm g}$ for all six polymers, see Table \ref{tab:tg}. 
Calculated $T_{\rm g}$ values in our simulations and its comparison to the experimental data is 
listed in the Table \ref{tab:tg}.
\begin{table}[ptb]
	\caption{Glass transition temperatures $T_{\rm g}$ of different polymers shown in Fig. \ref{fig:schem}.
       Available experimental data are also compiled for comparison.
       Note that we have obtained $T_{\rm g}$ value for PNIPAM in our experiments, 
       see supplementary material Section SI for more details.
}
\begin{center}
       \begin{tabular}{|c|c|c|c|c|c|c|c|c|c|c|c|}
\hline
Polymer         &   $T_{\rm g}$  (K)    &    $T_{\rm g}^{\rm exp}$  (K)  \\\hline
\hline
	       PMMA            &    390   &  378 \cite{polh}    \\
	       PAP             &    387   &  380 \cite{pipe14nm}   \\
	       PAA             &    400   &  385  \cite{pipe14nm}  \\
	       PAM             &    445   &  430  \cite{polh}  \\
PNIPAM          &    421   &  413    \\
P(NIPAM-co-AM)  &    455   &  $-$    \\
\hline
\hline
	       PS              &    $-$   &  373 \cite{polh}   \\
\hline                                              
\end{tabular}  \label{tab:tg}
\end{center}
\end{table}
Note that$-$ while the measured value of $T_{\rm g}$ for PNIPAM is obtained in our experiments, the 
other values are taken from the published literature \cite{pipe14nm,polh}. 

\begin{figure}[ptb]
\includegraphics[width=0.46\textwidth,angle=0]{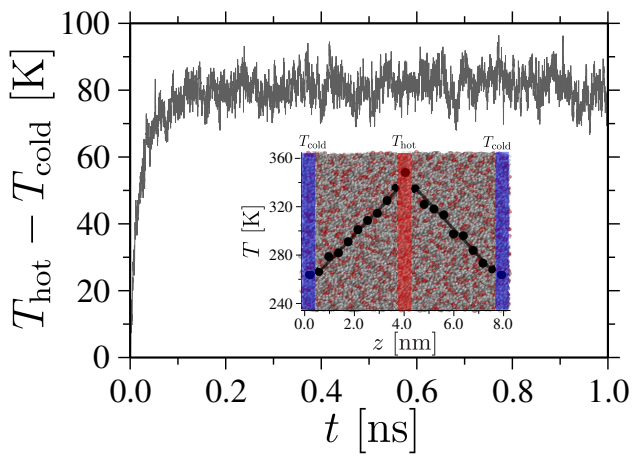}
	\caption{The main panel shows time equilibration of the temperature difference between the hot ($T_{\rm hot}$) and the cold ($T_{\rm cold}$) 
	regions of a solid PMMA material. In the inset we show the steady-state temperature profile along the $z-$axis.
	To establish the temperature gradient, the simulation domain is subdivided into 20 slabs with equal width. Within this setup, each 
	slab typically contains $\sim 3000$ atoms. 
\label{fig:mpmethod}}
\end{figure}

We employ the nonequilibrium method to calculate $\kappa$ \cite{plathe}. In this method, a heat flux $J$ through the material is generated 
in microcanonical ensemble by swapping atomic velocities between the hot $T_{\rm hot}$ and the cold $T_{\rm cold}$ region of the 
simulation box (see the inset of Fig. \ref{fig:mpmethod}). For this purpose the simulation cell is subdivided into 20 slabs along the $z-$axis.
Velocity swapping was performed between the slowest atom in the center slab (see the red region in the inset of  
Fig. \ref{fig:mpmethod}) and the fastest atom in the slab at the cell boundary (see the blue regions in the inset of  
Fig. \ref{fig:mpmethod}). This swapping was performed every 40 fs with a time step $\triangle t = 0.2$ fs. 
The system is initially equilibrated for 1 ns when the steady state of temperature difference between hot and cold regions 
is obtained, as shown in the main panel of Fig. \ref{fig:mpmethod}. After the steady state is reached, $J$ was then 
calculated for another 0.1 ns long simulation.
A linear fit of the temperature profile as a function of $z-$coordinate (see the inset of Fig. \ref{fig:mpmethod}) 
is used to calculate the thermal transport coefficient $\kappa$ using the Fourier's law of heat conduction, 
i.e., $\kappa = J/|\triangle T / \triangle z|$.

\subsection{Polymer synthesis}

To validate the ability of force field parameters of PNIPAM to reproduce properties of dry states, we have also synthesized 
PNIPAM sample using the reversible addition-fragmentation chain-transfer polymerization (RAFT) following the exact protocol 
discussed earlier for PNIPAM \cite{mukherji16sm}. $T_{\rm g}$ of this sample is measured using the differential 
scanning calorimetry (DSC) (see supplementary Fig. S2). As seen from the Table \ref{tab:tg}, measured and calculated
values of $T_{\rm g}$ for PNIPAM are in very good agreement. Note that$-$ given the fact that both PAM and PNIPAM
force fields are well validated, we expect to have reasonably captured the properties of solid copolymers with
these two constituents. Therefore, we abstain from synthesizing a P(NIPAM-co-AM) system.

\section{Results and discussions}
\label{sec:res}

\subsection{Thermal conductivity, elasticity and glass transition temperature}

\begin{figure}[h!]
\includegraphics[width=0.43\textwidth,angle=0]{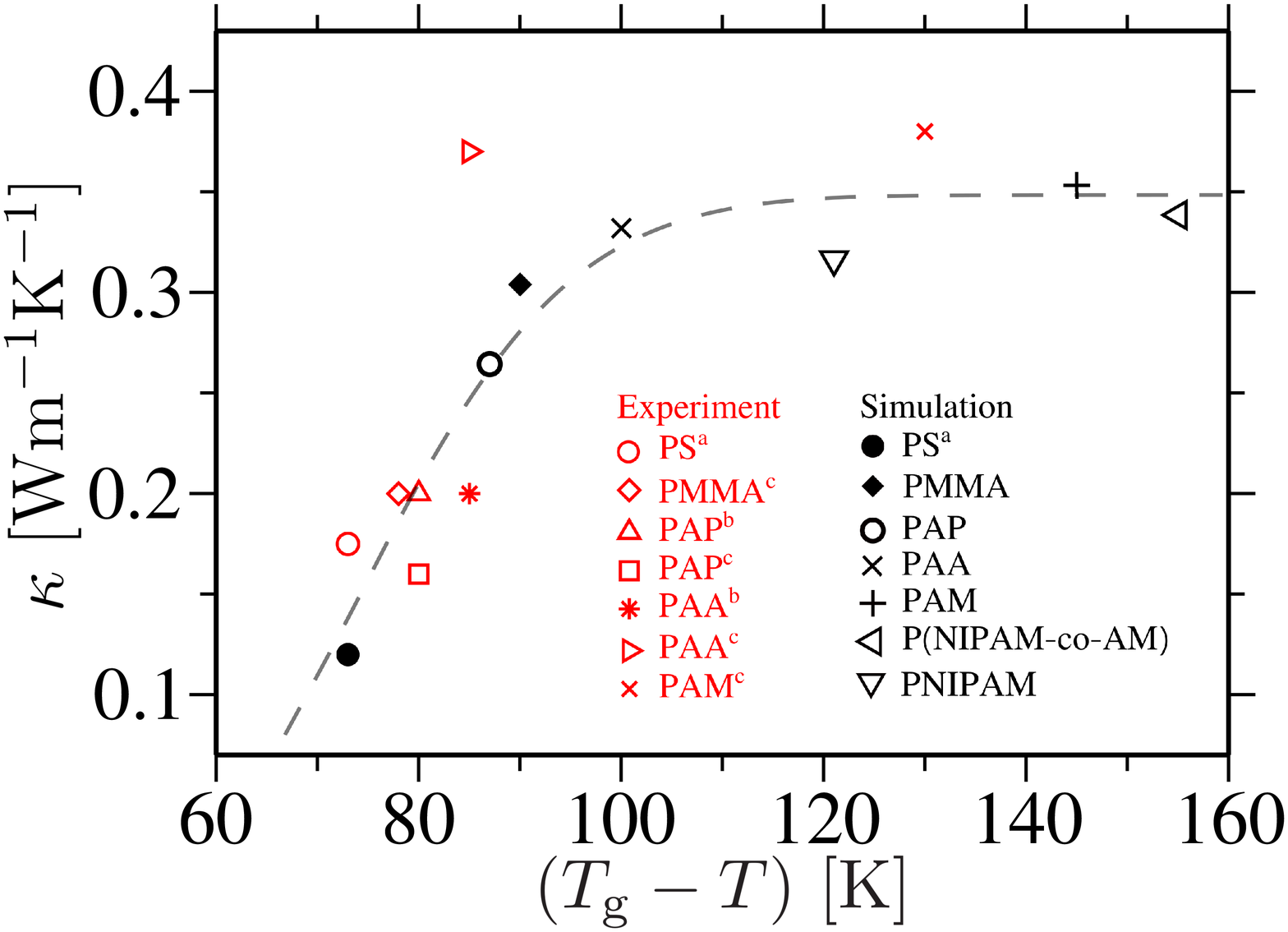}
	\caption{Thermal transport coefficient $\kappa$ as a function of the distance from the 
	 glass transition temperature $\left(T_{\rm g}-T\right)$. The data is shown for a temperature $T = 300$ K.
	 Calculated $\kappa$ values are compared with the available experimental data for several polymeric 
	 systems (as shown in Fig. \ref{fig:schem}). Data for $a$, $b$ and $c$ are taken from Refs. [\onlinecite{shenogin09jap}], 
	 [\onlinecite{pipe14nm}], and [\onlinecite{cahill16mac}], respectively. 
	 Typical errors of 5$-$10\% are estimated from four different $\kappa$ calculations for each 
	 polymeric system using different random seeds during microcanonical simulations. Line is drawn to guide the eye.
\label{fig:kappa_tg}}
\end{figure}

We start our discussion by investigating the effect of $T_{\rm g}$ on $\kappa$ for different solid polymers. 
In Fig. \ref{fig:kappa_tg} we show a comparative plot of $\kappa$ from our simulations and the corresponding 
experimental values from the published literature. 
It can be seen that $\kappa$ first increases up to $\left(T_{\rm g}-T\right) \simeq 100$ K,
i.e., dry polymers where the properties are dictated by weak vdW interactions. 
For $\left(T_{\rm g}-T\right) \geq 100$ K, $\kappa$ values remain constant with $T_{\rm g}$. 
These are H$-$bonded polymers. It can be seen that the overall trend in 
the variation of $\kappa$ for different polymers follows a reasonable master curve, 
with a couple of exceptions. For example, the experimental $\kappa$ values for PAA between two 
experiments \cite{pipe14nm,cahill16mac} deviate by a factor of two, while these individual 
experiments report typical error bars of about 5\%. In this context, we find that our 
calculated $\kappa$ for PAA is closer to the data reported in Ref. [\onlinecite{cahill16mac}].
However, our calculated $\kappa$ for PMMA is overestimated in comparison to experiments \cite{cahill16mac}.
It is also important to investigate why $\kappa$ remains invariant with $T_{\rm g}$ for H$-$bonded polymers and not for 
vdW-based systems? For this purpose, we need to look more closely into the microscopic interactions 
governing the glass forming behavior of these commodity plastics.

\begin{table}[h!]
	\caption{Average number of hydrogen bonds (H$-$bond) between non-bonded monomers for three 
	different polymeric systems.
	Data is shown for a temperature $T = 300$ K. For comparison we have also listed maximum number of possible
	H$-$bond between two non-bonded monomers.
}
\begin{center}
       \begin{tabular}{|c|c|c|c|c|c|c|c|c|c|c|c|}
\hline
Polymer         & H$-$bond  & Max H$-$bond  \\\hline
\hline
PAA             &    0.71     &  2.0    \\
PAM             &    1.41     &  4.0    \\
PNIPAM          &    0.51     &  2.0    \\
\hline
\end{tabular}  \label{tab:hbond}
\end{center}
\end{table}

To investigate the relationship between $T_{\rm g}$, microscopic interactions and $\kappa$, we have first 
calculated the average number of hydrogen bonds (H$-$bond) between non-bonded monomers. H$-$bond is calculated using the standard 
GROMACS subroutine, where a H$-$bond exists when the donor-acceptor distance is $\leq 0.35$ nm
and the acceptor-donor-hydrogen angle is $\leq 30^{\circ}$. In Table \ref{tab:hbond} we show H$-$bond for three different systems,
which span the full range $100 {\rm K}< \left(T_{\rm g}-T\right) < 160 {\rm K}$ where $\kappa$ remain invariant (see Fig. \ref{fig:kappa_tg}).
By comparing PAA and PAM data, we find that PAM has almost twice the number of H$-$bonds in comparison to 
PAA (see column two in Table \ref{tab:hbond}). This is not surprising given that 
PAM monomers have twice as many possibilities of forming H$-$bonds than PAA (see column three in Table \ref{tab:hbond} 
and corresponding monomer structures in Fig. \ref{fig:schem}). 
Furthermore, a higher number of H$-$bonds also leads to stronger interactions. For example, considering that the 
H$-$bond strength is between $4-8~k_{\rm B}T$, the energy per contact for PAM can be $\simeq 10~k_{\rm B}T$, 
while this is $\simeq 5~k_{\rm B}T$ for PAA. This trend is consistent with an almost 50 K increase in $T_{\rm g}$ 
for PAM in comparison to PAA (see Table \ref{tab:tg}).
Note that the contacts in these systems originate because of the interdigitation of the side groups, leading to
H$-$bonds.

\begin{figure}[ptb]
\includegraphics[width=0.49\textwidth,angle=0]{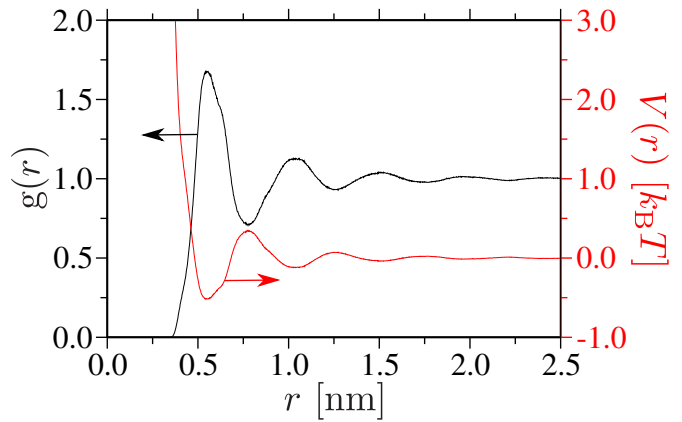}
	\caption{Radial distribution function ${\rm g}(r)$ and potential of mean force 
	$V(r) = -k_{\rm B}T\ln{\left[{\rm g}(r)\right]}$ for a PNIPAM system at $T = 300$ K. 
	Data is calculated between carbon atoms of the PNIPAM isopropyl group. 
	Arrows direct at the corresponding $y-$axes.
\label{fig:rdf}}
\end{figure}

It is not that H$-$bonds are always the dominant interactions in a system and vdW interactions can be 
completely ignored. Here PNIPAM is an interesting system$-$ while PNIPAM has less fewer H$-$bonds 
than PAA (see Table \ref{tab:hbond}), PNIPAM still has a higher $T_{\rm g}$ (see Table \ref{tab:tg}). 
This trend can be understood from the chemical structure of a NIPAM monomer 
that not only has the hydrophilic amide moiety, but also has a rather large hydrophobic isopropyl 
group (see PNIPAM structure in Fig. \ref{fig:schem}). 
Here the interaction between two carbon atoms of different isopropyl groups, belonging to 
two non-bonded monomers, can be estimated by using the Boltzmann inversion of 
the pair distribution function ${\rm g}(r)$. The carbon-carbon potential-of-mean-force can then 
be estimated using $V(r) = -k_{\rm B}T\ln{\left[{\rm g}(r)\right]}$ \cite{originbi1}.
It can be seen from the Fig. \ref{fig:rdf} that$-$ while two carbon atoms only interact by 
about $V(r) \simeq k_{\rm B}T/2$, collectively can lead to $>5~k_{\rm B}T$ interaction strength.
This is because one carbon atom in an isopropyl group can have as many as 8$-$10 neigbouring carbons.
Therefore, the vdW interactions between isopropyl groups also give almost similar 
contribution to the contact energy and thus to the $T_{\rm g}$ for PNIPAM.
Note that because of this delicate balance between vdW and H$-$bonded interactions in PNIPAM, we will
investigate this system in more details at a later stage of this manuscript.

The discussion in the preceding two paragraphs presents a nice consistency check between monomer structures, H$-$bond and $T_{\rm g}$.
However, it still remains counter-intuitive that $\kappa$ almost remains invariant (i.e., $\sim 0.32$ Wm$^{-1}$K$^{-1}$) 
even when $T_{\rm g}$ increases. Generally speaking, $T_{\rm g}$ is related to material stiffness, i.e., 
a material with higher $T_{\rm g}$ also exhibits higher stiffness. 
Therefore, it is worthwhile to investigate the relationship between the elastic response of different polymeric 
systems and its relationship with $\kappa$. In this context, it is well known that $\kappa$ is directly related to the elasticity of materials. 
A simple estimate for the lower bound of the thermal conductivity $\kappa^{\rm min}$, 
referred to as the minimum thermal conductivity model, relates to the sound wave velocities (or the materials stiffness) 
using the expression \cite{einstein,cahill92prb,cahill16mac},
\begin{equation}
	\kappa^{\rm min} = \left( \frac {\pi}{48} \right)^{1/3} k_{\rm B} n^{2/3} \left({v}_{\ell} + 2 {v}_{t} \right),
\label{eq:mintc}
\end{equation}
where $n$ is the atomic number density and $v_l = \sqrt{C_{11}/\rho}$ and $v_t = \sqrt{C_{44}/\rho}$ are the longitudinal 
and transverse sound wave velocities, respectively. Here, $C_{11} = K + 4 C_{44}/3$, $C_{44}$ is the shear modulus, $K$ is the 
bulk modulus and $\rho$ is the mass density. We use volume fluctuations to
calculate $K$ from NpT simulations using the expression,
\begin{equation}
        K = k_{\rm B}T \frac {\left<V\right>} {\left<V^2\right> - \left<V\right>^2}
	\label{eq:bulk}
\end{equation}
and this also leads to 
\begin{equation}
	C_{44} = \frac {3K \left( 1 - 2 \nu \right)} {2\left( 1 + \nu \right)},
	\label{eq:c44}
\end{equation}
with $\nu$ being the Poisson's ratio. $\nu$ is calculated from the isobaric uniaxial stretching of 
different samples, which was performed at a strain rate of $10^{-7}$ fs$^{-1}$ and at $T = 300$ K.
Details of the elastic constants and $\nu$ for different systems are listed in the 
supplementary Table SII. Note that Eqs. \ref{eq:bulk} and \ref{eq:c44} are good estimators when dealing with isotropic 
systems. For anisotropic systems, individual components of the elasticity tensor should be separately calculated. 

\begin{figure}[ptb]
\includegraphics[width=0.43\textwidth,angle=0]{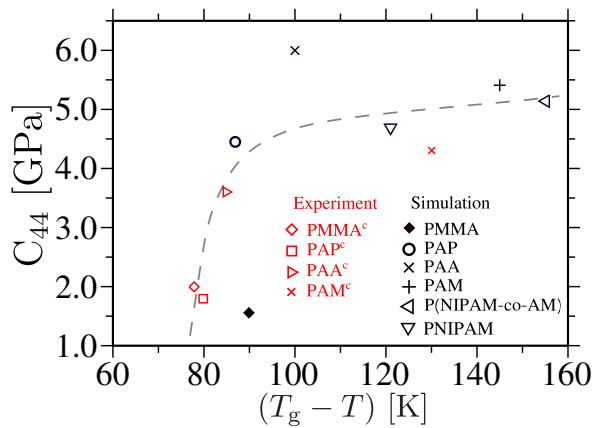}
	\caption{Shear modulus $C_{44}$ as a function of the distance from the glass transition temperature $\left(T_{\rm g}-T\right)$ 
	for different dry polymer systems. Data is shown for a temperature of $T = 300$ K. 
	Typical error of $\sim$20\% are estimated from four different calculations for each polymeric system.
	Experimental data $c$ are taken from Ref. [\onlinecite{cahill16mac}]. Line is drawn to guide the eye.
\label{fig:c44tg}}
\end{figure}

As predicted by Eq. \ref{eq:mintc}, $\kappa$ is related to $C_{44}$ and $C_{11}$. Therefore, we will now 
investigate the variation of elastic moduli for different systems. Note that$-$ while both $C_{44}$ and 
$C_{11}$ have very similar behavior with $T_{\rm g}$ for different systems (see supplementary Tables SII and SIII), 
here we only plot $C_{44}$ to explain the trend underlying microscopic picture. 
Therefore, in Fig. \ref{fig:c44tg} we show $C_{44}$ as a function of scaled $T_{\rm g}$.
Consistent with the trend in the Fig. \ref{fig:kappa_tg} we find that $C_{44}$ remains constant for $\left(T_{\rm g}-T\right) > 100$ K,
i.e., H$-$bonded systems. In this context, it is important to mention that all our systems are investigated
at least 80 K below their $T_{\rm g}$, such that their elastic moduli are
already in the glassy plateau of the sigmoidal curve. Therefore, these systems can be treated within the 
harmonic approximation. Since the mechanical response is dictated by 
small (local) particle fluctuations, the corresponding elastic constants are then given by the curvature of 
the potential energy surface around the equilibrium particle positions. 
Ideally speaking, the high dimensional potential energy surface of a macromolecular amorphous 
system is extremely complex. Microscopically, however, this complex energy surface can be
decomposed into  different contributions. In the case when H$-$bonds play significant role in dictating properties of commodity plastics,
the stiffness is related to the particle fluctuations around their typical H$-$bond length of 0.34 nm.
Furthermore, a H$-$bond is inherently directional and restrictively local in nature.
Note that locality of a H$-$bond originate because, when an oxygen atom forms a bond with a 
hydrogen, the same hydrogen atom involved in this pair can not simultaneously interact 
with another oxygen via H$-$bond. Furthermore, because of the dense packing of side chains, even individual 
oxygen atoms can only form a maximum of one H$-$bond. While these bonds dynamically break and form, instantaneously 
these are strictly restricted between a pair of particles. Therefore, so long as interactions 
are H$-$bond dominated, materials stiffness remains invariant irrespective of the specific monomer structure.

When the microscopic interactions are predominantly vdW$-$based,
such as PMMA, PAP or PS, each atom experiences an effective interaction due to the presence 
of all other particles within the first neigbouring shell. This effectively depends on the monomer 
architecture and their chemical constituents, thus can alter the contact energy density, $T_{\rm g}$ and 
also the stiffness, see data for $\left(T_{\rm g}-T\right) < 100$ K in Fig \ref{fig:c44tg}. 
However, the plateaus observed in $\kappa$ and $C_{44}$ for H$-$bonded systems 
within the range 100 K$<\left(T_{\rm g}-T\right)<$ 160 K (see Figs. \ref{fig:kappa_tg} and \ref{fig:c44tg}) also suggest 
that there exists a restrictive range of $\kappa$ for the H$-$bonded plastics, i.e., 0.3 Wm$^{-1}$K$^{-1}$ $< \kappa \leq0.4$ Wm$^{-1}$K$^{-1}$.
Furthermore, this maximum in $\kappa$ is independent of specific chemical structure of a monomer
and $T_{\rm g}$. 

\subsection{Thermal conductivity, elasticity and minimum thermal conductivity model}

We will now compare our simulation data with the minimum thermal conductivity model presented 
in Eq. \ref{eq:mintc} \cite{cahill92prb}. For this purpose, we have calculated $v_{\ell}$ and $v_t$ in our simulations.
These values and corresponding elastic moduli are presented in the supplementary Tables SII and SIII. 
\begin{figure}[ptb]
\includegraphics[width=0.46\textwidth,angle=0]{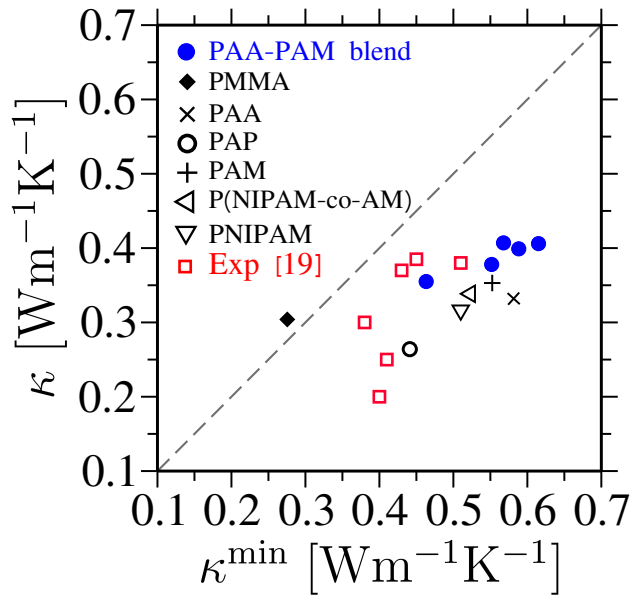}
	\caption{Calculated thermal transport coefficients $\kappa$ as a function of $\kappa^{\rm min}$ obtained
	using Eq. \ref{eq:mintc}. Dashed line is a linear law with zero intercept and unity slope.
	Data is shown for dry states of six different linear polymer systems and also for the PAM-PAA blends \cite{mukherji19mac}.
	For comparison, we have also included data from the published experimental literature \cite{cahill16mac}.
	Note that we have used $n_c = 2(n-n_{\rm H})/3$ instead of $n$ in Eq. \ref{eq:mintc} 
	with $n_{\rm H}$ being the density of H$-$atoms, see the text for more details.
\label{fig:mintc}}
\end{figure}
In Fig. \ref{fig:mintc} we show $\kappa$ from simulations as a function of $\kappa^{\rm min}$.
Data is shown for six different polymers investigated in this study, reanalysis of our earlier published 
simulation work of H$-$bonded polymer blends \cite{mukherji19mac} and the available experimental data \cite{cahill16mac}. 
It can be seen that $\kappa$ and $\kappa^{\rm min}$ values show a reasonable correlation. 
Furthermore, a slight overestimation of $\kappa^{\rm min}$ in all cases is due to the fact that we are dealing with rather complex molecular architectures,
which is not effectively accounted in Eq. \ref{eq:mintc}. 
Something that speaks in this favor is that$-$ atomic solids, without any long 
range chain-like connectivity, show better agreement between $\kappa$ and $\kappa^{\rm min}$ \cite{cahill92prb}.
Additionally, the data in Fig. \ref{fig:mintc} further suggests that $\kappa$ is indeed bound by
an upper limit of $\sim 0.4$ W/m$^{-1}$K$^{-1}$ for hydrogen bonded systems resulting from the 
upper bound of elasticity. 

It should also be mentioned that all calculations are performed at $k_{\rm B}T$, 
which is too small to excite high frequency modes associated with the hydrogen atoms attached
to the carbon atoms of different monomers and also the chemically bonded interactions within a chain
that often exceed $80k_{\rm B}T$ strength. If all these modes are included in $n$ of Eq. \ref{eq:mintc}, 
it will lead to an overestimation of $\kappa$.
This can, however, be overcome following the treatment proposed in Refs. [\onlinecite{cahill17prb,cahill16mac}]
that replaces $n$ by $n_c = 2(n-n_{\rm H})/3$ in Eq. \ref{eq:mintc}, 
with $n_{\rm H}$ being the density of H$-$atoms. This also gives an upper classical (Dulong-Petit) 
limit of the volumetric heat capacity $c_{\rm DP} = 3n_ck_{\rm B}$, and following Eq. \ref{eq:mintc} gives,
$	\kappa^{\rm min} = \left( {\pi}/{432} \right)^{1/3} k_{\rm B}^{1/3} c^{2/3} \left({v}_{\ell} + 2 {v}_{t} \right)$ \cite{cahill92prb,cahill16mac}.
In our simulations, we have estimated $c$ using the scaled atomic density $n_c$ and
from the energy fluctuation in canonical ensemble, i.e., $c_f = \left[\left<E^2\right> - \left<E\right>^2\right]/Vk_{\rm B}T^2$. 
In Table \ref{tab:cv} we list $c$ for four different polymers where experimental data is available.
\begin{table}[ptb]
\caption{Volumetric heat capacity $c$ for poly(methyl methacrylate) (PMMA), poly({\it N}-acryloyl piperidine) (PAP),
       poly(acrylic acid) (PAA), and poly(acrylamide) (PAM). In simulations, heat capacity is calculated using the energy
	fluctuations $c_f$ and also the Dulong-Petit (DP) values $c_{\rm DP} = 3n_ck_{\rm B}$. 
	For comparisons, we have also listed available experimental values $c_{\rm exp}$ \cite{cahill16mac}.
}
\begin{center}
       \begin{tabular}{|c|c|c|c|c|c|c|c|c|c|c|c|}
\hline
	       Polymer         &   $c_f$ (MJ/m$^{3}$K)   &    $c_{\rm DP}$ (MJ/m$^{3}$K) &    $c_{\rm exp}$ (MJ/m$^{3}$K) \\\hline
\hline
               PMMA            &    1.40  &   1.22  & 1.65 \\
	       PAP             &    1.01  &   1.23  & 1.67 \\
               PAA             &    1.25  &   1.84  & 1.49 \\
	       PAM             &    1.50  &   1.68  & 1.67  \\
\hline
\end{tabular}  \label{tab:cv}
\end{center}
\end{table}
Reasonably good agreement between $c_f$, $c_{\rm DP}$, and $c_{\rm exp}$ further suggests that our simulations capture the generic 
features of $\kappa$ of commodity plastics.

\subsection{Effect of chain stretching on thermal conductivity}

As predicted by Eq. \ref{eq:mintc}, if the stiffness of a material is increased, one also expects to 
increase $\kappa$. This points to a more general design principle of tunability in polymeric plastics. 
In this work, when we specify an upper bound of $\kappa$, we limit our discussion to the standard 
commodity plastics that are dictated either by the vdW or the H$-$bond interactions. 
For zwiterionic polymers or for the ionic systems, where long-range electrostatic interactions 
play significant role, this upper bound can be increased to a value 
of $\kappa \sim 0.6$ W/m$^{-1}$K$^{-1}$ \cite{cahill17prb}. 
$\kappa$ can be even improved further by electrostatic engineering \cite{kim17sci,luo19jpcc}. 

Other studies have also shown that $\kappa$ can be significantly improved by using stretched polymer
configurations that is governed by the phonon-like vibrational excitations that are along the covalently 
bonded polymer backbone. For example, a large body of works have been conducted on 
fiber-like polymer configurations to enhance $\kappa$ \cite{chen08prl,yang12prb,shen10natnano}.

\begin{figure}[ptb]
\includegraphics[width=0.49\textwidth,angle=0]{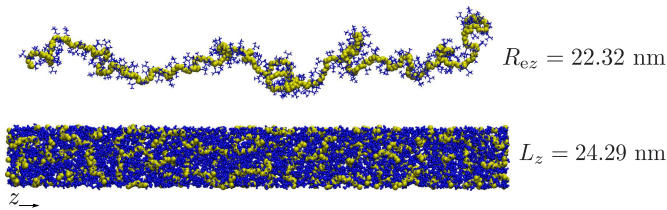}
        \caption{Simulation snapshot of a fiber like configuration of dry
        poly({\it N}-isopropyl acrylamide) (PNIPAM) sample with a chain length $N_{\ell} = 256$,
        which corresponds to 100 persistence lengths $\ell_p$. Top panel shows a
        single chain configuration within the simulation box spanning across the simulation box.
        Bottom panel shows typical simulation box consisting of four elongated chains.
        Yellow spheres represent the alkane backbone and the side chains are highlighted in blue.
\label{fig:pni256}}
\end{figure}

To investigate the effect of polymer stretching, we have made use of a PNIPAM chain with $N_{\ell} = 256 = 100~\ell_p$.
This chain was previously equilibrated in (good solvent) pure methanol \cite{mukherji16sm}. 
A typical configuration of an expanded (solvent free) single chain oriented along the $z-$axis is shown in the 
top panel of Fig. \ref{fig:pni256}.
A simulation sample is prepared by placing four expanded chains in a box. This box has an equilibrium
$z-$dimension of $L_z = 24.29$ nm at $T = 300$ K, which is comparable to the $z-$component of the 
end-to-end distance $R_{ez} = 22.32$ nm (see lower panel of Fig. \ref{fig:pni256}).
For this system we find $\kappa = 0.70$ Wm$^{-1}$K$^{-1}$. i.e., a two fold increase in $\kappa$ with 
respect to the bulk value of PNIPAM (see supplementary Table SI). 
We have also calculated Young's modulus $E \sim 10$ GPa of the elongated sample. This value is very similar 
to the bulk data for PNIPAM (see the supplementary Table SII). 
The bulk like elasticity of this configuration is 
not surprising given that$-$ even when the individual chains are rather elongated, $R_{ez}$ is only about 
1/3 of the total contour length $\ell_c \simeq 60.0$ nm of a PNIPAM chain with $N_{\ell} = 256$. 
Therefore, elasticity is not influenced by the bond stretching, rather only by the vdW and H$-$bond 
based non-bonded interactions in a PNIPAM chain.

A closer look at a PNIPAM chain shows that PNIPAM consists of a hydrophobic alkane (PE like) backbone 
with hydrophilic side (amide) groups. If this chain is fully stretched along its contour, 
such that $R_{ez} \simeq \ell_c$, it should also show similar $\kappa$ values as in the case of single 
stretched PE. Here, for a PE chain with $N_{\ell} = 256$, $\kappa \simeq 25$ Wm$^{-1}$K$^{-1}$ \cite{yang12prb}
that is almost 40 fold increase from the $\kappa$ observed for our stretched sample shown in Fig. \ref{fig:pni256}.
The increase of $\kappa$ for stretched configuration is also coupled with the significant increase in carbon-carbon
bond stretching modulus, which is about $E > 250$ GPa \cite{pe1}.

\section{Conclusion}
\label{sec:conc}

Using large scale molecular dynamics simulations of all-atom polymer models we investigate thermal transport of 
commodity plastics and its links to the mechanical response. For this purpose, we have simulated 
six different linear polymeric materials and one polymer blend with five different mixing ratios \cite{mukherji19mac}.
We have investigated the effect of increasing interaction strength, i.e., going from vdW to H$-$bond
\cite{deju}, on the material stiffness and $\kappa$ of solid polymeric materials.
Our simulation data suggest that there is an upper limit to the achievable $\kappa$ of commodity plastics, 
so long as the systems are restricted to van der Waals and hydrogen bonding (H$-$bond) based plastics. 
This is because the local and directional nature of H$-$bonds limit the maximum materials stiffness between 4.5-5.5 GPa, 
as measured by the shear modulus $C_{44}$. This upper limit also limits the highest achievable 
$\kappa \simeq 0.40$ Wm$^{-1}$K$^{-1}$. Specific chemical structure and the glass transition temperature $T_{\rm g}$
are found to play no role in dictating tunability of $\kappa$ of commodity plastics. 
We have also investigated the effect of covalent bonds \cite{deju} on $\kappa$ by simulation of a
chain oriented polymer configuration. The results presented here are consistent with the minimum thermal conductivity 
model \cite{cahill92prb} and existing experimental data \cite{cahill16mac}. 
While this work correlates microscopic molecular coordination with elasticity and $\kappa$, 
the concepts presented here may pave ways for the better tunability of the physical properties of the 
common and smart polymeric materials. Therefore, we expect this to have far reaching implications in 
designing environmental friendly materials for advanced functional uses.\\

\noindent{\bf Acknowledgement:}
D.M. thanks Martin H. M\"user, Alireza Nojeh, Daniel Bruns, and Tiago Espinosa de Oliveira 
for many useful discussions. We further acknowledge support from Compute Canada where 
simulations were performed. J.R. thanks the Alexander von Humboldt Foundation for financial support.

\end{document}